\documentclass[aps,prb,twocolumn]{revtex4-1}
\usepackage{gensymb}
\usepackage{graphicx}
\usepackage{array}
\newcolumntype{P}[1]{>{\centering\arraybackslash}p{#1}}
\usepackage{amsmath}

\begin{document}
\title{Structural and electronic properties of Fe(Al$_x$Ga$_{1-x}$)$_3$ system}
\author{Debashis Mondal$^{1,2}$, C. Kamal$^1$, Soma Banik$^1$, Ashok Bhakar$^1$, Ajay Kak$^3$, Gangadhar Das$^{1,2}$, V. R. Reddy$^{4}$, Aparna Chakrabarti$^{1,2}$, Tapas Ganguli$^{1,2}$}
\email{tapas@rrcat.gov.in}
\address{$^1$Indus Synchrotrons Utilization Division, Raja Ramanna Centre for Advanced Technology, Indore, 452013, India.}
\address{$^2$Homi Bhaba National Institute, Training School Complex, Anushakti Nagar, Mumbai, 400094, India.}
\address{$^3$Mechanical $\&$ Optical Support Section, Raja Ramanna Centre for Advanced Technology, Indore-452013, India.}
\address{$^4$UGC-DAE Consortium for Scientific Research, University Campus, Khandwa Road, Indore, 452001,India.}
\begin{abstract}
FeGa$_3$ is a well known d-p hybridization induced intermetallic bandgap semiconductor. In this work, we present the experimental and theoretical results on the effect of Al substitution in FeGa$_3$, obtained by x- ray diffraction (XRD), temperature dependent resistance measurement, room temperature Mossbauer measurements and density functional theory based electronic structure calculations. It is observed that upto $x$ = 0.178 in Fe(Al$_x$Ga$_{1-x}$)$_3$, which is the maximum range studied in this work, Al substitution reduces the lattice parameters $a$ and $c$ preserving the parent tetragonal P4$_2$/mnm crystal structure of FeGa$_3$. The bandgap of Fe(Al$_x$Ga$_{1-x}$)$_3$ for $x$ = 0.178 is reduced by $\approx$ 24\% as compared to FeGa$_3$. Rietveld refinement of the XRD data shows that the Al atoms replace Ga atoms located at the 8j sites in  FeGa$_3$. A comparison of the trends of the lattice parameters and energy bandgap observed in the calculations and the experiments also confirms that Al primarily replaces the Ga atoms in the 8j site.
\end{abstract}
\maketitle
\section{Introduction}
Intermetallic alloys and compounds of transition metal elements are an interesting category of materials with a large number of technological applications because of their exotic mechanical, electronic and magnetic properties \cite{Stoloff}. Although most of the intermetallic alloys and compounds are conductors, there are quite a few transition metal based intermetallic stoichiometric compounds such as FeSi, FeSb$_2$, RuAl$_2$, RuGa$_3$, FeGa$_3$ and RuIn$_3$ \cite{Schlesinge,Petrovic,Weinert,Amagai,Bogdanov}, where an energy gap at the Fermi level (E$_f$)  has been experimentally observed. In some systems like FeAl$_2$, RuAl$_2$, RuGa$_2$ and OsGa$_2$, where experimental data have not been reported, first-principle calculations have predicted a semiconducting gap \cite{Mihalkovic}. Among the transition metal based intermetallics, FeGa$_3$ is one of the most studied systems due to its potential application as a thermoelectric material and also from a fundamental point of view. Theoritical studies on FeGa$_3$, and its d$^6$-III analogs like RuGa$_3$, OsGa$_3$ and RuIn$_3$ systems have been reported and they all stabilize in the P4$_2$/mnm structure with a bandgap of 0.50 eV, 0.26 eV, 0.68 eV and 0.30 eV respectively \cite{Imai}. In all these systems, the origin of bandgap has been attributed to the hybridization between the d orbitals of the transition element and the p orbital of the group III element. These low bandgap semiconductors with large density of states near the Fermi level show a promise for application as thermoelectric material. FeGa$_3$ shows a large negative Seebeck coefficient with the value of 563 $\mu$V/K for polycrystalline sample\cite{Hadano} and 350 $\mu$V/K for single crystal \cite{Umeo} which has attracted the deep interest for an understanding of the electronic structure of this material.

The bandgap of FeGa$_3$ has been experimentally determined by various ways. Temperature dependent resistivity measurements have led to a gap of $\approx$ 0.20 eV \cite{Haldolaarachchige}, 0.26 eV \cite{Amagai} and 0.50 eV \cite{Hadano,Umeo}. Temperature dependent magnetic succeptibility measurement at high temperature shows an activated type behavior, with the bandgap values between 0.29-0.45 eV \cite{Tsujii}. More recently, Arita et al. have determined the bandgap to be 0.4 eV using a combination of photoelectron spectroscopy (PES) and inverse photoelectron spectroscopy (IPES) experiments \cite{Arita}. First-principles based electronic structure calculations with local density approximation (LDA) predict its bandgap to be around 0.3-0.5 eV \cite{Haussermann,Imai}, which is in good agreement with the experimental values.

Modification of the physical properties brought about by substitution at various sites in these types of intermetallic compound, have also been studied quite extensively. Substitution of the Fe site with Co and Ga sites with Ge improves the figure of merit for thermoelectric applications  by a factor of five as compared to FeGa$_3$ \cite{Haldolaarachchige}. The results of a study on the effect of gradual substitution of Ge in FeGa$_{3-y}$Ge$_y$ shows that a little amount of Ge substitution (upto $y$ = 0.006), results in metallic conduction \cite{Umeo} and as the amount of Ge substitution is increased ($y$ = 0.13), a weak ferromagnetic order is introduced. Co and Ge doping on the Fe and Ga sites respectively in FeGa$_3$ give an extra electron, and thereby reduce the resistivity and improves the thermoelectric properties.

All the above studies on the substitution of elements in FeGa$_3$, correspond to ones with a different electronic configuration as compared to Fe and Ga. The effect of these substitution on the magnetic and thermoelectric  properties have been reported. But, there are no reports on detailed studies on the isoelectronic element substitution in FeGa$_3$. There has been only one report on an isoelectronic substitution (Al) in Ga site, where samples were synthesized by spark plasma sintering process \cite{Wagner}. The reported  thermoelectric property of the 0.02\% Al substituted FeGa$_3$, was nearly the same as the parent compound. In the present work we have substituted Al  in FeGa$_3$ upto 17.8\% and studied the changes in the lattice parameters, bandgap and electronic structure with this substitution. These studies are important from the point of applications to bandgap engineering in FeGa$_3$ based IR devices. We have observed a decrease in the lattice parameters $a$ and $c$ and the bandgap, with an increase in Al substitution. Furthermore, first-principles calculations have also been carried out to understand the observed bandgap variation with increase in Al concentration.

\section{Experiment}
All samples have been prepared in an induction furnace using high purity elements (Fe = 99.98\%, Ga = 99.999\%, Al = 99.999\%) in 99.999\% pure Ar gas atmosphere. Elements were stacked in a few layers, inside a graphite crucible for homogeneous melting. The induction melting frequency was 15 kHz.  The samples were melted twice for homogeneous melting. After melting, the samples were annealed for 5 days at $600^{\circ}$C inside a quartz ampoule, which was vacuum sealed at  $2\times10^{-6}$ mbar pressure.  Synchrotron based energy dispersive x-ray fluorescence (XRF) was used for the determination of the elemental compositions of all the annealed samples.

\begin{figure}
 \centering
 \includegraphics[width=8 cm, height=5.0 cm]{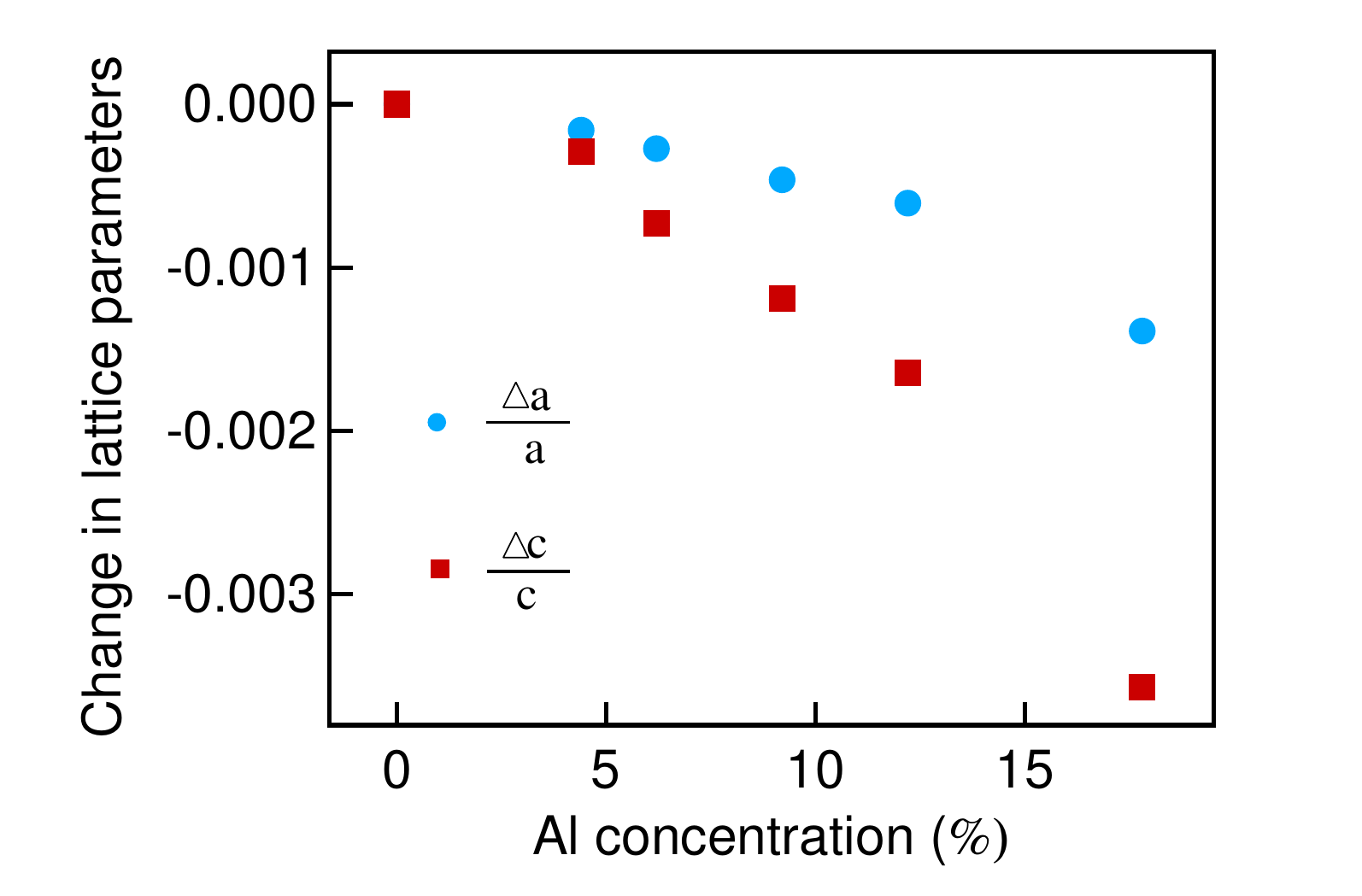}
 \caption{Change in lattice parameters with Al concentration in Fe(Al$_x$Ga$_{1-x}$)$_3$.}
 \end{figure}

X-ray diffraction (XRD) measurements have been made on a Bruckers D8 Advance powder diffraction system using  Cu $K_\alpha$ radiation. Finely powdered sample was rotated at a speed of $\approx$ 20 rpm during the diffraction measurement. The structures of all the samples and their lattice parameters have been determined from XRD measurements and the detailed structural analysis have been carried out using Rietveld analysis. Temperature dependent resistance measurements in the range of 300 K to 750 K have been made on all the samples inside a 99.999\% pure Ar gas purged horizontal furnace. A Kiethley 6220 current source and Kiethley 2182A nanovoltmeter have been used for the resistance measurements. The slope of ln R vs. 1/T (Arrhenious plots) have been used to determine the activation energy E$_a$ using the relation R = R$_o$exp(-E$_a$/2kT). For low bandgap materials like FeGa$_3$ measurements carried out in the 300 K to 750 K range correspond to the intrinsic conduction region. Thus the evaluated activation energy corresponds to the bandgap of the material \cite{Tsujii}.

\begin{table*}
\centering
 \begin{tabular}{P{1cm}P{1cm}P{1.5cm}P{1.5cm}P{1.5cm}P{1.2cm}P{1.2cm}P{1.4cm}}
\hline
Sample No&	$Al\%$ $\pm$1 &	$a = b$ (\AA) $\pm$0.0001&	$c$ (\AA) $\pm$0.0001&	Bandgap (eV) $\pm$0.01&	FWHM (mm/s) $\pm$ 0.01&		Isomar shift (mm/s) $\pm$ 0.01&	Quardpole splitting (mm/s) $\pm$ 0.01\\
\hline
1&	0.0&		6.2669&		6.5602&		0.46&			0.37 &	0.26 &		0.32 \\
2&	4.4&	6.2667&		6.5590&		0.43&					&				&						\\
3&	6.2&	6.2652&		6.5554&		0.42&			0.38 &	0.27 &		0.32 \\
4&	9.2&	6.2640&		6.5524&		0.42&					&				&						\\
5&	12.2&	6.2631&		6.5494&		0.40&			0.35 &	0.26 &		0.31 \\
6&	17.8&	6.2582&		6.5368&		0.35&			0.36 &	0.23 &		0.29 \\
\hline
\end{tabular}
\caption{Variation of lattice parameters and bandgap with Al substitution for all the samples along with the Mossbauer results for four samples (0\%, 6.2\%, 12.2\% and 17.8\% Al substituted).}
\end{table*}

Room temperature Mossbauer spectra have been recorded for four different samples to check the presence of local magnetic ordering. Experiments have been performed on powdered samples in transmission mode with a $^{57}$Co (Rh) radioactive source in constant acceleration mode using a standard PC-based Mossbauer spectrometer equipped with a WissEl velocity drive. Velocity calibration of the spectrometer was carried out with a natural iron absorber at room temperature.

\begin{figure}
\centering
\includegraphics[width=9 cm, height=12 cm]{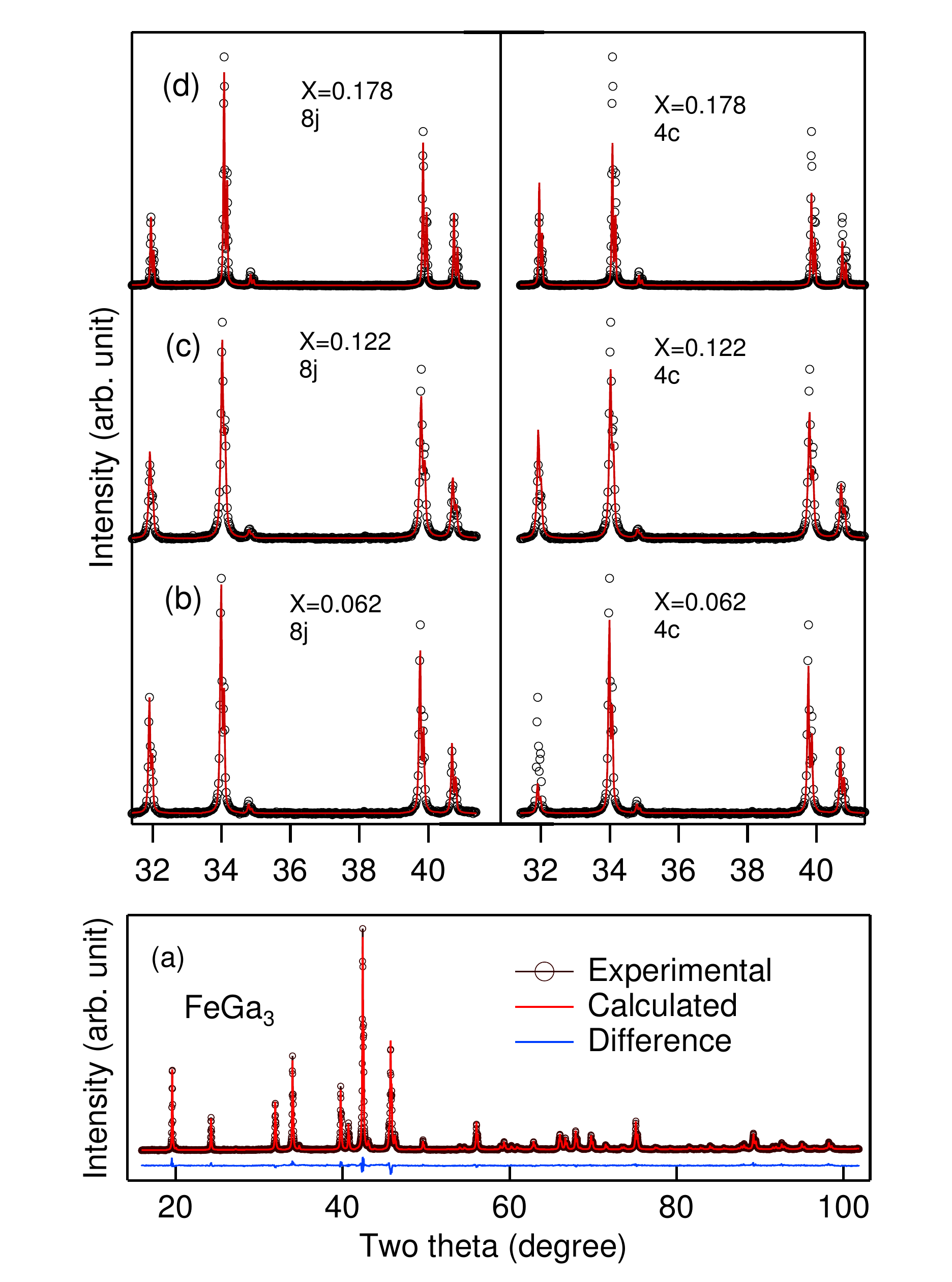}
\caption{XRD patterns of different samples, using Cu $K_\alpha$ source and their Rietveld refined pattern are shown. (a) XRD pattern of FeGa$_3$ ranging from 2$\theta$ = 14.2$^\circ$ to 103.2$^\circ$, (b-d)  highlighted portion of XRD patterns of Fe(Al$_x$Ga$_{1-x}$)$_3$ for $x$ = 0.062, 0.122 and 0.178 ranging from 2$\theta$ = 31.4$^\circ$ to 41.4$^\circ$ with Al occupying 4c and 8j sites for each case.}
\end{figure}

In order to understand and explain the experimental data, we have also performed density functional theory (DFT) based calculations using the Vienna ab-initio simulation package VASP \cite{Kresse1,Kresse2} within the framework of the projector augmented wave (PAW) method. For the exchange-correlation potential,  we have employed the generalized gradient approximation (GGA) over local density approximation given by Perdew, Burke, and Ernzerhof (PBE) \cite{Perdew}. The cutoff for the plane wave expansion has been taken to be 400 eV. The mesh of k-points for Brillouin zone integration is chosen to be 11 $\times$ 11 $\times$ 10. The convergence for the plane wave cutoff and the number of k-points in the mesh have been checked by varying these parameters. The convergence criterion for energy in SCF cycles is chosen to be 10$^{-6}$ eV. All the structures are optimized by minimizing the forces on individual atoms with the criterion that the total force on each atom is below 10$^{-2}$ eV/\AA. The calculation was carried out on FeGa$_3$ (four formula units in the unit cell) and Fe$_4$Ga$_{11}$Al, which corresponds to 0\% and 8.33\% Al substitution ($x$ = 0.0833) in Fe(Al$_x$Ga$_{1-x}$)$_3$ respectively.

\section{Result and discussion}
Table I gives the Al composition in all the samples that have been determined by XRF measurements. The XRD data of all the samples correspond to the reported P4$_2$/mnm structure data in literature  (PDF$\#$ 04-010-4423). The XRD data of all these samples were analyzed by Rietveld refinement and the values of $a$ and $c$, determined for these samples are also given in Table I. A continious decrease in the lattice parameters $a$ and $c$ is observed in the samples with an increase in the Al concentration within the range studied in this work. It is further observed that with Al incorporation upto 17.8\%, the change in $a$ is 0.14\% whereas the change in $c$ is 0.36\%. We have plotted $\frac{\Delta a}{a}$ and $\frac{\Delta c}{c}$ with respect to the substituted Al percentage in Fig 1. It clearly shows the larger change in lattice parameter $c$, compared to the lattice parameter $a$. 

\begin{figure}
\centering
\includegraphics[width=9 cm, height=4.5 cm]{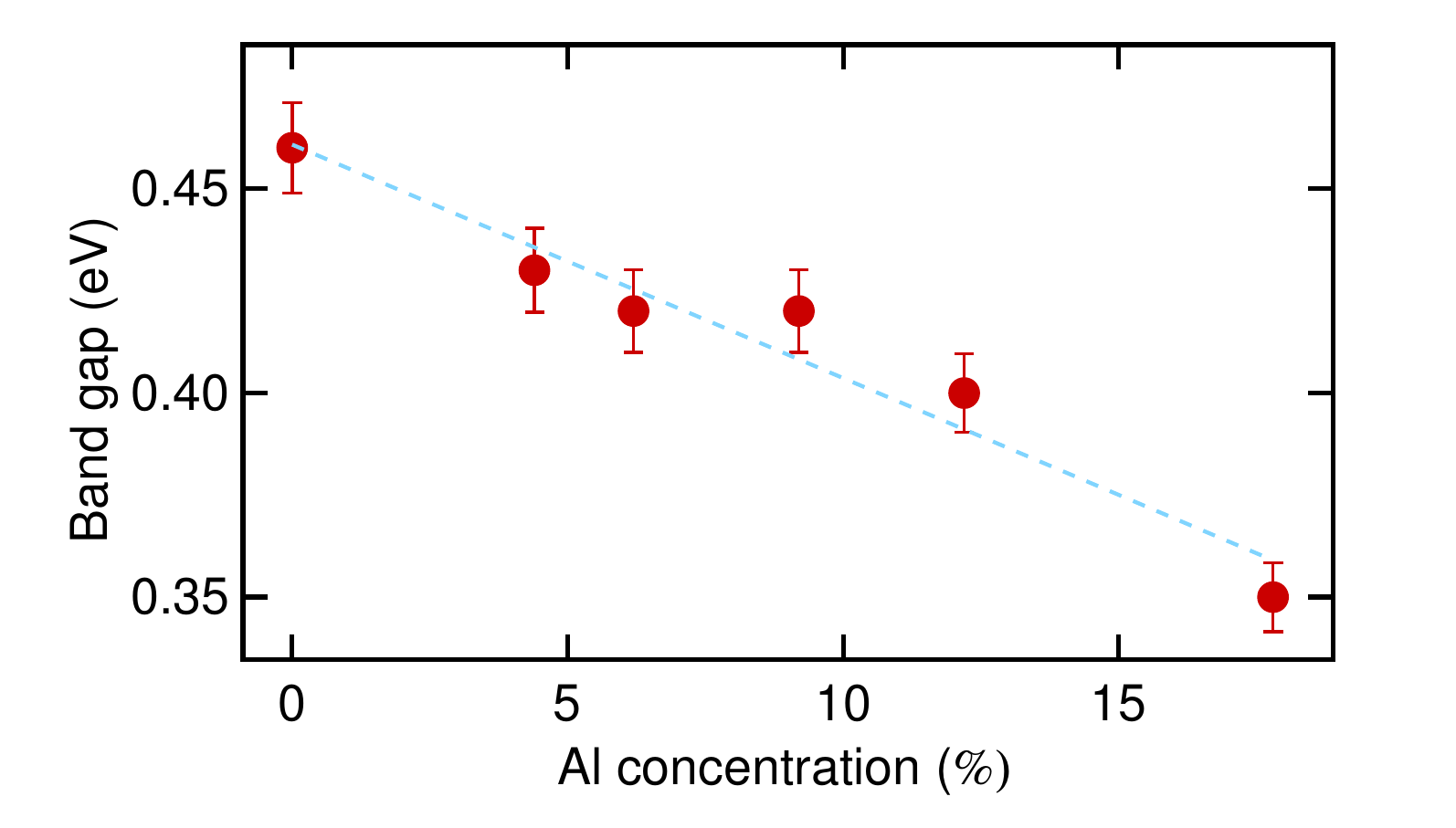}
\caption{Bandgap variation with Al substitution. Markers represent the actual data and the dotted line is a guide to eye.}
\end{figure}

The Rietveld refinement of the XRD data of FeGa$_3$ is shown in Fig 2(a). Good fitting is obtained for the XRD data and the simulated pattern of FeGa$_3$ with P4$_2$/mnm symmetry. The obtained structural parameters and Wyckoff positions of Fe and Ga atoms (mentioned in Table-II) match very well with the values available in the literature \cite{Haussermann}. The Fe atoms occupy the 4f site and the Ga atoms are found to have two different symmetry positions, 4c site (0.344, 0.344, 0.000), which is referred to as Ga1 site and 8j site (0.156, 0.156, 0.262), which is referred to as Ga2 site. In the case of Al substituted samples, there is a possibility that Al atoms can occupy both the 4c or 8j site.

Figure 2 (b to d) also shows the graphs of the XRD data along with the Rietveld refinement results for samples with $x$ = 0.062, 0.122, and 0.178, in the 2$\theta$ region ranging from 31.4$^\circ$ to 41.4$^\circ$. The refined data with Al substituting the Ga atoms in the 4c and the 8j sites are indicated for all the three samples separately. This particular 2$\theta$ region is selected for presentation, so that the significant misfit corresponding to the Al substitution at the 4c site is highlighted. These results show that throughout the range of $x$ studied in this work, the refinement is better for Al occupying the 8j site. The results obtained from ab-initio calculations also show that the formation energy of Fe(Al$_x$Ga$_{1-x}$)$_3$ with Al occupying the 8j site is lower as compared to Al occupying the 4c site (shown in Table III). Thus, 8j site is more favourable for Al compared to 4c site, which supports our XRD results. We thus conclude that, Al atoms substitute the Ga atoms in 8j (Ga2) site and not in the 4c (Ga1) sites. The results of Rietveld refinement for Fe(Al$_x$Ga$_{1-x}$)$_3$ ($x$ = 0.178) is shown in Table II.

\begin{center}
\begin{table*}
 \begin{tabular}{P{1cm}P{1.8cm}P{1.8cm}P{1.8cm}P{1.8cm}P{1cm}}
\hline
$Atom$ &	$Position$ &		$x$ &	$y$ &	$z$ &	$occ$\\
\hline
		&				& FeGa$_3$&			&		&			\\
\hline
Fe&			4f&		$0.34482(24)$&	$0.34482(24)$&	$0.00000$&	$1.00$\\
Ga1&		4c&		$0.00000$&	$0.50000$&	$0.00000$&	$1.00$\\
Ga2&		8j&		$0.15616(12)$&	$0.15616(12)$&	$0.26217(17)$&	$2.00$\\
\hline
		&				& Fe(Al$_x$Ga$_{1-x}$)$_3$&			&		&			\\
\hline
Fe&			4f&		$0.34293(21)$&$	0.34293(21)$&	$0.00000$&	$1.00$\\
Ga1&		4c&		$0.00000$&	$0.50000$&	$0.00000$&	$1.00$\\
Ga2&		8j&		$0.15638(14)$&	$0.15638( 14)$&	$0.26310( 18)$&	$1.46$\\
Al&			" &						"		&			"				&			"				&	$0.54$\\
\hline
\end{tabular}
\caption{Rielveled refinement result for FeGa$_3$ and Fe(Ga$_{0.822}$Al$_{0.0178}$)$_3$. For Fe(Ga$_{0.822}$Al$_{0.0178}$)$_3$ sample, Al are spread randomly in 8j sites.}
\end{table*}
\end{center}

The bandgap determined from the Arrhenius plot for all the samples is plotted in Fig. 3. The markers represent the experimental data and the dotted line is a guide to eye. The results show that the bandgap decreases  with increasing concentration of Al in Fe(Al$_x$Ga$_{1-x}$)$_3$. There is a reduction in the bandgap from 0.46 eV to 0.35 eV, which is a reduction of $\approx$ 24\% with a substitution of $\approx$ 17.8\% of Al in FeGa$_3$.

\begin{figure}
\centering
\includegraphics[width=5 cm, height=8 cm]{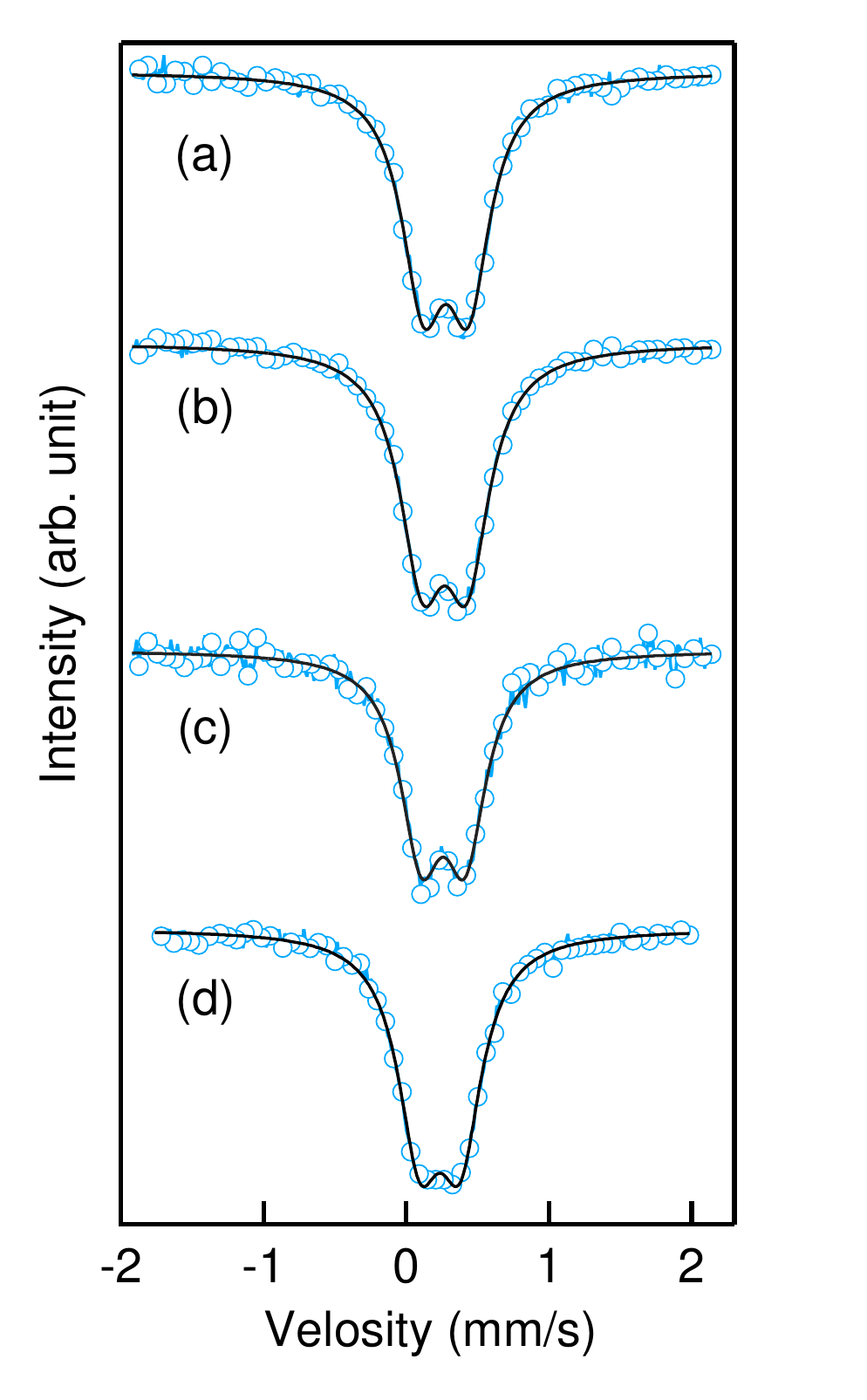}
\caption{Room temperature Mossbauer spectra of Fe(Al$_x$Ga$_{1-x}$)$_3$ with (a) x = 0, (b) x = 0.062, (c) x = 0.122 and (d) x = 0.178.}
\end{figure}

Room temperature Mossbauer spectra of all the samples are shown in Fig. 4 and the results of this experiment are summarized in Table I.  FeGa$_3$ Mossbauer data of the present work exhibit similar trend as found in earlier reports \cite{Tsujii, Dezsi}. Further, with Al doping, the spectra remained a quadrupole-split doublet. Width of the lines (FWHM), isomer shift and quadrupole splitting for all the samples remain same, within the experimental error. This confirms that Al substitution does not change the local neighbourhood of Fe i.e., it does not influence the bonding and the hybridization in Fe(Al$_x$Ga$_{1-x}$)$_3$ significantly. The observed doublet with Al substitution, indicates that there is no change in the local internal magnetic field at the Fe site.

To understand the observed reduction in the bandgap with increasing Al substitution, we discuss the results of the first principles calculations. Calculation of Fe$_4$Ga$_{11}$Al (corresponding to $x$ = 0.0833) with Al occupying the 8j site shows a small decrease in the bandgap of $\approx$ 0.01 eV (change by 2.3\%). A reduction in the lattice parameters $a$ (by 0.06\%) and $c$ (by 0.14\%) is observed with Al substitution. In contrast to this, when Al occupies the 4c site, larger reduction in the bandgap of $\approx$ 0.1 eV (change by 22.2\%) is observed. However, there is an increase in the lattice parameter $a$ ( by 0.29\%) with increase in Al concentration, which is contrary to our experimental observation. Furthermore, it is observed from the results of formation energy that when Al is substituted in 8j site the system is energetically more favourable as compared to the case of 4c site. These results of the calculations are summarized in Table III. If we interpolate the experimental data for the lattice parameter and the bandgap for Al concentration of $x$ = 0.0833, we find that the change in the bandgap and the lattice parameters $a$ and $c$ are 8.7\%, 0.05\% and 0.11\% respectively. The values of the variations in the lattice parameters for Al substituting the 8j site, obtained from experiments and calculations are very close. The experimental values of the bandgap variation with Al substitution is larger than the calculated values for 8j site substitution. This relatively larger decrease in the bandgap obtained from experiments as compared to the calculations is possibly due to the following reasons: (a) a finite probability of some Al atoms occupying the 4j site (it may be noted that there is a large decrease in the bandgap with Al substituting the 4c site), (b) presence of unintentional defects in the form of vacancies and interstitials etc resulting in local stress and hence a reduction of the bandgap. The reduction in the bandgap with stress has been observed in FeGa$_3$ in our experimental work on high pressure studies and will be reported in a separate communication. Reduction in the band gap with pressure has also been reported recently in a theoretical calculation \cite{Osorio}.

\begin{table*}
\begin{tabular}{P{3cm} P{1.5cm}P{1.5cm} P{1.3cm}P{1.5cm}P{1.5cm}P{1.6cm}}
\hline
System&		Lattice Constant $a$ ({\AA})&		Lattice Constant $c$ ({\AA})&		Total Energy (eV)&	Formation energy (kJ/mol)&	Magnetic moment ($\mu_B$)&	Bandgap (eV)\\
\hline
FeGa$_3$&										6.267&	6.540&	-73.391&	-91.990&		0.000&	0.43\\
Fe$_4$Ga$_{11}$Al (Al in 8j) &		6.263&	6.531&	-74.243&	-381.360&	0.000&	0.42\\
Fe$_4$Ga$_{11}$Al (Al in 4c) &		6.285&	6.521&	-74.144&	-371.760&	0.000&	0.34\\
\hline
\end{tabular}
\caption{DFT based first principle calculation results are summarized for FeGa$_3$ and FeGa$_{11}$Al with Al sitting at 8j and 4c site.}
\end{table*}

Figure 5 shows the total density of states (DOS) and the partial density of states of the Fe, Ga and Al atoms as evaluated from first principles calculations. Since our results from experiments and the corresponding calculations indicate that the Al atoms mainly occupy 8j site, the DOS for this case  only is shown. We find that the contribution to the density of states per atom to the DOS from the Al atom specifically close to E$_f$ is less than the contribution from Ga atom. This may be related to the reduced size of the substituted Al atom in place of the Ga atom. The atomic radius of Ga is 136 pm which is $\approx$ 15\% larger than the atomic radius of Al which is 118 pm \cite{Periodic}; Thus the d-p orbital overlap between  the Fe and  Al orbitals is expected to be significantly reduced, which might result in a reduced bandgap on Al substitution.

\begin{figure}
 \centering
 \includegraphics[width=9 cm, height=12 cm]{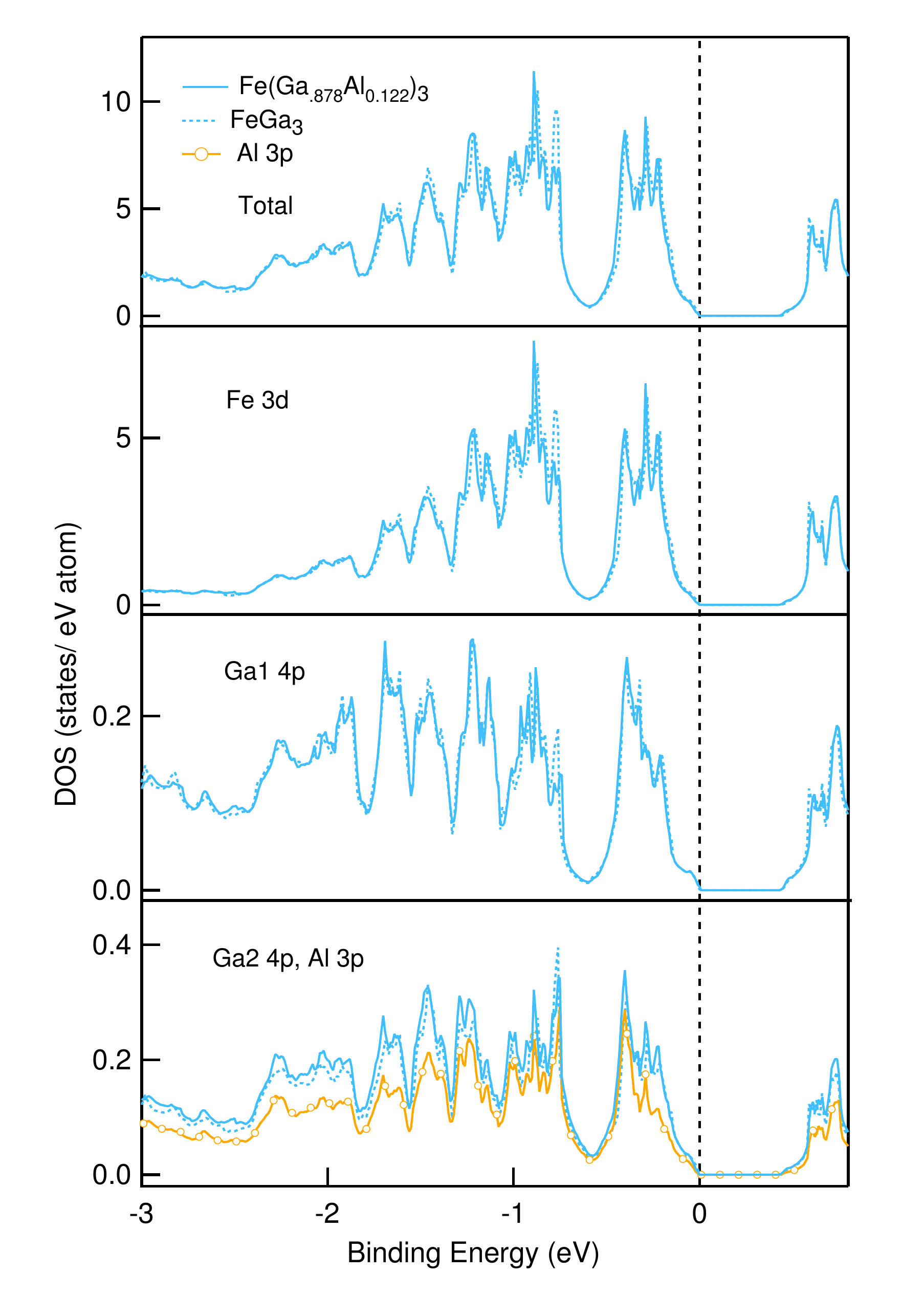}
 \caption{Total and partial DOS obtained from DFT based ab-initio calculation for FeGa$_3$ and Fe(Ga$_{11}$Al)$_3$. Al 3p DOS are represented by yellow colored dashed line.}
 \end{figure}

The origin of larger reduction in bandgap with Al substituting the 4c (Ga1) site compared to 8j (Ga2) site can be understood from the crystal structure of FeGa$_3$. As Ga1 and Fe atom lie in the same XY plane, 3d$_{xy}$ orbital of Fe atom will strongly overlap with 4s orbital of Ga1 atom and will partially overlap with 4p$_x$ and 4p$_y$ orbital of Ga1 atom. In addition, 3d$_{x^2-y^2}$ orbital of Fe atom will also strongly overlap with 4p$_x$ and 4p$_y$ orbital of Ga1 atom. In contrast to this, the Ga2 and Fe atoms do not lie in same XY, YZ and ZX plane, and thus only 3d$_{xy}$ orbital of Fe atom will overlap with 4p$_z$ orbital of Ga2 atom; 3d$_{yz}$ orbital of Fe atom will overlap with 4p$_x$ orbital of Ga2 atom; and d$_{zx}$ orbital of Fe atom will overlap with 4p$_y$ of Ga2 atom. Further, the distances between Fe and Ga2 atoms are 2.399 {\AA} (near) and 2.498 {\AA} (far), whereas the distance between Fe and Ga1 atoms are 2.370 {\AA}. So the Al atom at 4c site interact strongly with Fe atom compared to the Al atom occupying the 8j site. Thus, the hybridization induced bandgap is primarily dominated by the Fe-Ga1 orbital overlap and the Fe-Ga2 orbital overlap contribute relatively less to the bandgap.

The observation of a reduction in the bandgap with increasing Al substitution is interesting because in all III-V semiconductors an opposite trend is observed in the bandgap variation with Al substituting the Ga site. For example, on the substitution of Al in GaAs, the bandgap increases by about 12\% for 12\% Al substitution \cite{Vurgaftman}. In GaP, the bandgap increases by $\approx$ 3\% with 12\% Al substitution and in GaSb, the bandgap increases by 23\% with 12\% Al substitution. This observation is primarily explained in terms of the bandgaps of the two end compounds, where in all the cases, the Al containing compound has a much higher bandgap as compared to their corresponding Ga containing counterpart. The bandgap in all such III-V compounds is due to the energy difference between the p bands of the group III and V elements at the top of the valence band and the s bands of the group III and V elements at the bottom of the conduction band. The difference in the energy of s and p orbital is higher in the case of Al based compounds, because of its lower atomic number, as compared to the corresponding Ga based compounds, and hence the obsrved increase in the bandgap with Al substitution in III-V systems. In contrast, in  FeGa$_3$, the bandgap is due to hybridization beween d and p orbitals of Fe and Ga atoms respectively. As a result, when the smaller sized Al atom is substituted in place of relatively larger sized Ga atoms, the overlap of d-p orbitals and hence hybridization strength is reduced. This causes a reduction in the bandgap. This reduction in the bandgap with Al substitution in FeGa$_3$ can have an important consequence for the development of IR devices based on bandgap engineering.

\section{Conclusion}
In this work, we have determined the changes in lattice parameter and the bandgap of  Fe(Al$_x$Ga$_{1-x}$)$_3$  for 0$\leq x \leq$  0.178. We find that lattice parameters ($a$ and $c$) decrease with increase of Al concentration and change in lattice parameter $c$ is more than the change in the parameter $a$. The smaller sized Al atom replaces the Ga atom in the 8j site. There is also a decrease in the bandgap of Fe(Al$_x$Ga$_{1-x}$)$_3$ with an increase in $x$. These present results provide a promising way to engineer the bandgap in this relatively new low bandgap material with Al substitution for infrared device applications.

\section{Acknowledgement}
Authors would like to thank Madhusmita Baral for her support in sample preparation, Uday Deshpande and Himal Bhatt for IR measurements. C. Kamal and A. Chakrabarti would like to thank Computer Division , RRCAT for their technical support. Authors also would like to thank RRCAT for financial support.

 \end{document}